\documentstyle[12pt,epsfig]{article}
\newlength{\dinwidth}
\newlength{\dinmargin}
\newcommand{\ba}{\begin{array}}
\newcommand{\ea}{\end{array}}
\newcommand{\beq}{\begin{equation}}
\newcommand{\eeq}{\end{equation}}
\newcommand{\bea}{\begin{eqnarray}}
\newcommand{\eea}{\end{eqnarray}}

\def\ep{\varepsilon}

\def\S{{\bf S}}
\def\C{{\bf C}}
\def\bce{\begin{center}}
\def\ece{\end{center}}

\def\nonu{\nonumber}

\def\pa{\partial}
\def\al{\alpha}
\def\be{\beta}
\def\ga{\gamma}
\def\Ga{\Gamma}

\def\De{\Delta}
\def\ep{\epsilon}

\def\th{\theta}

\def\la{\lambda}
\def\La{\Lambda}

\def\ph{\phi}

\def\ps{\psi}

\def\R{{\bf R}}
\def\Z{{\bf Z}}
\def\RP{{\bf RP}}
\def\S{{\bf S}}

\begin{document}
\thispagestyle{empty}
\addtocounter{page}{-1}
\begin{flushright}
{\tt hep-th/9908162}\\
\end{flushright}
\vspace*{1.3cm}
\centerline{\Large \bf ${\cal N}=2$ SCFT and M Theory
on $AdS_4 \times Q^{1,1,1}$}
\vspace*{1.5cm} 
\centerline{ \sc Changhyun Ahn
}
\vspace*{1.0cm}
\centerline{\it 
Department of Physics,}
\vskip0.3cm 
\centerline{  \it Kyungpook National University,}
\vskip0.3cm
\centerline{  \it  Taegu 702-701, Korea }
\vspace*{0.3cm}
\centerline{\tt ahn@kyungpook.ac.kr 
}
\vskip2cm
\centerline{\large\bf \sc Abstract}
\vspace*{0.5cm}

Coincident M2 branes at a conical singularity are related to
M theory on $AdS_4 \times X_7$ for an appropriate 7 dimensional 
Sasaki-Einstein manifold $X_7$. For $X_7=Q^{1,1,1}=(SU(2) \times SU(2) \times
SU(2))/(U(1) \times U(1))$ which was found sometime ago, 
the infrared limit of the theory on $N$ M2 branes
was constructed recently. It is 
the $SU(N) \times SU(N) \times SU(N)$ gauge theories
with three series of chiral fields $A_i, i=1,2$ transforming in the 
$\bf (N, \overline{N},1)$ representation, $B_j, j=1,2$ transforming in the 
$\bf (1,N, \overline{N})$ representation and $C_k, k=1,2$
transforming in the $\bf (\overline{N},1,N)$ representation.
From the scalar Laplacian of $X_7$ on the supergravity side, 
we discuss the spectrum of
chiral primary operators of dual ${\cal N}=2$ superconformal field 
theory in 3 dimensions.
We study M5 branes wrapped over 5-cycle of $X_7$ which were 
identified as (three types of) 
baryon like operators made out of $N$ chiral fields
recently. We consider 
M5 brane wrapped over 3-cycle of $X_7$ which plays the role of
domain wall in $AdS_4$. 
The new aspect arises when baryon like operators(M5 branes wrapped over
5-cycle) cross a domain wall(M5 brane wrapped over 3-cycle), 
M2 brane between them must be created. 

\vspace*{4.0cm}

\begin{flushleft}
{Aug., 1999}\\
\end{flushleft}
\baselineskip=18pt
\newpage

\section{Introduction}

In \cite{mal}  the large $N$ limit of superconformal field theories (SCFT) 
was described by taking the
supergravity limit on anti-de Sitter (AdS) space.
The scaling dimensions of operators of SCFT can be obtained from the 
masses of particles in string/M theory \cite{polyakov,wi}. In particular, 
${\cal N}=4$ $SU(N)$ super Yang-Mills theory in 4 dimensions is described by
Type IIB string theory on $AdS_5 \times {\bf S}^5$.  
This AdS/CFT correspondence was tested by studying
the Kaluza-Klein (KK) states of supergravity theory and 
by comparing them with the chiral primary operators
of the SCFT on the boundary. 
There exist also ${\cal N} =2, 1, 0$ superconformal theories in 
4 dimensions which have corresponding supergravity description 
on orbifolds of $AdS_5 \times {\bf S}^5$.
The field theory/M theory duality also provides
a supergravity description on $AdS_4$ or $AdS_7$
for some superconformal theories in 3 or 6 dimensions. 
The maximally supersymmetric theories and
the lower supersymmetric cases were also realized on the worldvolume of 
M theory at orbifold singularities. 

The gauge group of the boundary theory becomes 
$SO(N)/Sp(2N)$ \cite{witten1} by taking
appropriate orientifold operations for 
the string theory on $AdS_5 \times \S^5$. 
According to general arguments in \cite{witten1}, Type IIB string theory
on $AdS_5 \times X_5$ where $X_5$ is a five dimensional Einstein manifold
with five-form flux is dual to a four dimensional SCFT.
In \cite{kw} it was found that for the $X_5=(SU(2) \times SU(2))/U(1)$, the 
string theory on $AdS_5 \times X_5$ can be described by ${\cal N}=1$ 
supersymmetric $SU(N) \times SU(N)$ gauge theories coupled to four 
bifundamental chiral superfields and supplemented by a quartic 
superpotential.   
A field theory analysis of anomalous three point funtion reproduced
\cite{gubser} the central charge expected by supergravity. 
Baryon like chiral operators \cite{gk} made out of $N$ chiral superfields 
were identified with D3 branes wrapped over the 3-cycle of $X_5$ and domain
wall in $AdS_5$ was interpreted as D5 brane wrapped over 2-cycle of $X_5$.
Moreover the full KK spectrum analysis was done in a series of paper 
\cite{ceresoleetal,ceresoleetal1}. 

By generalizing the work of \cite{witten1} to the
case of $AdS_7 \times \RP^4$ where the eleventh dimensional
circle is one of $AdS_7$ coordinates, $(0,2)$ six dimensional SCFT
on a circle rather than uncompactified full M theory was described in
\cite{aky}.
For $SU(N)$ $(0,2)$ theory, a wrapped D4 brane on $\S^4$ together
with fundamental strings was
interpreted as baryon vertex. Furthermore 3 dimensional extension 
\cite{akly} was obtained by considering D6 branes wrapping on $\RP^6$.  
Backgrounds of the form $AdS_4 \times X_7$ arise as the near horizon
geometry of a collection of M2 branes in M theory \cite{mp}. 
Many examples where
$X_7$ is a coset manifold $G/H$ were studied in the old days of KK theories.
It is natural to ask {\it 
what is dual superconformal field theory corresponding to
M theory on  $AdS_4 \times X_7$?} 
As a first step, we will consider only
$X_7=Q^{1,1,1}$ in this paper. 
Recently the dual theory corresponding to 
this specific compactification in \cite{fabbrietal} turns out to be 
a nontrivial infrared
fixed point. It is the $SU(N) \times SU(N) \times SU(N)$ gauge theories
with three series of chiral fields $A_i, i=1,2$ transforming in the 
$\bf (N, \overline{N},1)$ representation, $B_j, j=1,2$
 transforming in the 
$\bf (1,N, \overline{N})$ representation and 
 $C_k, k=1,2$
 transforming in the 
$\bf (\overline{N},1,N)$ representation.
\footnote{A different dual SCFT proposal was made in \cite{ot,dall}.} 
The global symmetry of the gauge theory is $SU(2) \times SU(2) \times SU(2)$
where each of doublets of chiral fields transforms in the fundamental 
representation of one of the $SU(2)$'s.

Since $Q^{1,1,1}$  described as the coset spaces $G/H$ where $G=SU(2)\times
SU(2) \times SU(2)$ and $H=U(1) \times U(1)$
has the isometry $SU(2) \times SU(2) \times SU(2) 
\times U(1) $ \cite{dfv}, 
we are looking for an isolated singularity Calabi-Yau
fourfold with this symmetry. That is, 
the geometry of eight dimensional cone is Calabi-Yau fourfold while that of
seven dimensional $X_7$ is Sasaki-Einstein manifold \cite{mp,afhs}.
\footnote{
The minimal supersymmetric ${\cal N}=1$ CFT analysis and its RG flow have 
been studied when M2 branes are located at the conical singularity on eight 
dimensional manifold with $Spin(7)$ holonomy in \cite{ar}.}  
Let us consider the complex manifold $C$,
\bea
z_0^2+z_1^2+z_2^2 +z_3^2+
z_4^2+z_5^2+z_6^2+z_7^2 =0
\label{Cmani}
\eea
on $\C^8$.
This equation describes a surface which is smooth apart from the
origin.  The apex or node is a double point, i.e. a singularity for which
$C=0$ and $d C=0$ but for which the matrix of second derivatives is
nondegenerate. Note that if $z_i$ solves (\ref{Cmani}) so does
$\la z_i$ for any $\la$, so the surface
is made of complex lines through the origin and is a cone. 
The base of the cone is given by the intersection of the space
of solution of (\ref{Cmani}) with a sphere of radius 1 in $\C^8$,
\bea
|z_0|^2+|z_1|^2+|z_2|^2 +|z_3|^2+
|z_4|^2+|z_5|^2+|z_6|^2+|z_7|^2 =1.
\eea
These eight coordinates, invariant $\C^{*}$ action described by
toric geometry \cite{dall} can be expressed by triple product of six chiral
fields $A_i, B_j, C_k$ with some independent embedding equations.
It turns out that 
two D term equations modded out by the action of two $U(1)$'s give exactly
the manifold $(\S^3 \times \S^3 \times \S^3)/(U(1) \times U(1))$.   
The metric on the conifold may be written as
\bea
ds^2  = dr^2
 +r^2 g_{ij} dx^i dx^j, \;\;\;\; (i,j=1,2, \cdots, 7)
\eea
where $g_{ij}$ is the metric on the base of the cone that is exactly
$Q^{1,1,1}$. The radial coordinate $r$ is identified with the fourth
coordinate of $AdS_4$ and the section of the cone is identified with
the internal manifold $Q^{1,1,1}$. See also many papers 
\cite{uranga,dm,gns,lopez,unge,dm1,kw1}
dealt with conifold
singularity.

In this paper, in section 2, we recapitulate the spectrum of 
scalar Laplacian on $Q^{1,1,1}$ found in \cite{pope} sometime ago,
using the metric and seven coordiantes of $Q^{1,1,1}$ explicitly.
The hypermultiplet spectrum in KK harmonic expansion on $Q^{1,1,1}$ agrees 
with the chiral fields predicted by dual conformal field theory as shown by
\cite{fabbrietal}. In section 3, we clarify 
the property of three baryon like operators identified as M5 branes 
wrapped around 5-cycle. In section 4, we claim that
M5 brane wrapped over 3-cycle of $Q^{1,1,1}$ plays the role of
domain wall in $AdS_4$ and explain corresponding dual field theory
when the baryon like operators cross the domain wall. Finally we will come to
remaining wrapped branes.  

\section{Laplacian of $Q^{1,1,1}$ and Spectrum of Chiral Operators}

To describe the spectrum of chiral primaries in the $Q^{n_1, n_2, n_3}$ CFT
we need the expression for scalar Laplacian on $Q^{n_1, n_2, n_3}$.
Since
$X_7=Q^{n_1, n_2, n_3}$ is a $U(1)$ bundle over $\S^2 \times \S^2 \times \S^2$,
we take the spherical polar coordinates $(\th_i, \ph_i), i=1, 2, 3$ 
to parametrize $i$-th two sphere, as usual, and the angle 
$\ps$ parametrizes the $U(1)$ fiber.
By inverse Kaluza-Klein method,
the seven dimensional metric consists of $U(1)$ fiber coordinate together
with a potential and six dimensional base $\S^2 \times \S^2 \times \S^2$.
From the most general expression for harmonic two-form $U(1)$ field strength, 
the metric on $Q^{n_1, n_2, n_3}$ is given by \cite{np,pp,pp1}
\bea
 g_{ij} dx^i dx^j = 
c^2 \left( d\ps +\sum_{i=1}^{3} n_i \cos \th_i d \ph_i \right)^2
 +\sum_{i=1}^{3} \frac{1}{\La_i} \left( d \th_i^2+\sin^2 \th_i d \ph_i^2
\right)
\label{metric1}
\eea
where $c$ is a constant, $\La_i$ are scaling factors and 
$n_i$'s characterize the winding numbers of the $U(1)$ field over three 
$\S^2$'s.
The angles vary over the ranges
\bea
{\th}_i \in (0, \pi), \;\;\; {\ph}_i \in (0, 2\pi), \;\;\; 
\ps \in (0, 4\pi).
\eea 
All $n_i$'s must be integers because we assigned period $4 \pi$ on the
variable $\psi$.  Notice that all three integers $n_i$'s are necessary to
characterize these spaces(while $M^{pqr}$ space can be characterized by
only two integers).
One obtains inverse metric $g^{i,j}$ from (\ref{metric1}) and the nonzero 
components
are
\bea
g^{0,0}& = & \frac{1}{c^2} +\sum_{i=1}^3 \La_i n_i^2 \mbox{cot}^2 \th_i,
\;\;\; g^{2i-1,2i-1}= \La_i,
\;\;\; g^{2i, 2i} = \La_i \mbox{csc}^2 \th_i, \nonu \\
g^{0,2i} & = & g^{2i,0}= -\La_i n_i \mbox{cot} \th_i \mbox{csc} \th_i,
\;\;\; i=1, 2, 3.
\eea
We get the volume of $Q^{n_1, n_2, n_3}$ by integrating $\sqrt{g}=
\sqrt{\mbox{det} g_{ij}}= c \; \mbox{sin} \th_1 \mbox{sin} 
\th_2 \mbox{sin} \th_3/(\La_1
\La_2 \La_3)$ over 
the allowed range of variables,
\bea
\mbox{Vol} \left( Q^{n_1, n_2, n_3} \right) 
= 256 \pi^4 \frac{c}{\La_1 \La_2 \La_3}.
\label{volumeQ}
\eea

By using the seven coordinates, determinant $g$  
and inverse metric components explicitly,
the Laplacian can be expressed as
\bea
\Box \Phi  & = & \frac{1}{\sqrt{g}} \frac{\pa}{\pa x^i} 
g^{ij} \sqrt{g} \frac{\pa}{\pa x^j} \Phi
 \nonu \\
&  
= & \left( 
\frac{1}{c^2} \frac{\pa^2}{\pa \ps^2} +\sum_{i=1}^3 \La_i 
\left( n_i  \mbox{cot} \th_i
\frac{\pa}{\pa \ps} -  \mbox{csc} \th_i 
 \frac{\pa}{\pa \ph_i} \right)^2 
+\sum_{i=1}^3 \frac{1}{\mbox{sin} \th_i} \frac{\pa}{\pa \th_i}   \La_i
\mbox{sin} \th_i \frac{\pa}{\pa \th_i} \right) \Phi \nonu \\
& = & -E \Phi.
\eea
This can be solved by separation of variables eventhough the $Q^{n_1, n_2,
n_3}$ is not a product space. By writing
\bea
\Phi=\left( \prod_{i=1}^3 \Phi_i(\th_i) \right)  \mbox{exp} \left( i 
\sum_{i=1}^3 m_i
\phi_i \right) \mbox{exp} \left( i s \psi \right),
\eea
we get
\bea
E= \sum_{i=1}^3 \La_i E_i +\frac{s^2}{c^2}
\label{eigen}
\eea
where $E_i$'s satisfy the ordinary differential equations
\bea
\left( \frac{1}{\mbox{sin} \th_i } \frac{\pa}{\pa \th_i}  \mbox{sin} \th_i
\frac{\pa}{\pa \th_i} - \left(  s n_i \mbox{cot}
\th_i - m_i \mbox{csc} \th_i \right)^2 \right) \Phi_i =-E_i \Phi_i.
\label{diff}
\eea
Notice that for $n_i=0$ and $n_i=1$ 
(\ref{diff}) determines the eigenvalues of the
Laplacian on $\S^2$ and $\S^3$ respectively.
By defining $z_i =\mbox{cos}^2 \frac{\th_i}{2}$, it is easy to see
that (\ref{diff}) becomes
a hypergeometric equation with a solution
\bea
\Phi_i= z_i^I (1-z_i)^J F(A, B, C; z_i)
\eea
where $A,B,C,I,J$ are smooth in the interval ${\th}_i \in (0, \pi)$
and have a behavior
at the end points that can be determined by usual formula \footnote{
It is well known in any mathematical tables that $
F(A,B;C;z_i)= \frac{\Ga(C) \Ga(C-A-B)}{\Ga(C-A) \Ga(C-B)} F(A,B;A+B-C+1;1-z_i)+
(1-z_i)^{C-A-B} \frac{\Ga(C) \Ga(A+B-C)}{\Ga(A) \Ga(B)} 
F(C-A,C-B;C-A-B+1;1-z_i).$}.
The solutions are regular when they can be expressed in terms of 
a hypergeometric polynomials. It is known that \footnote{
It turns out that $A=\frac{1}{2}+m_i -\sqrt{\frac{1}{4}+E_i +|s n_i|^2},
B= \frac{1}{2}+m_i +\sqrt{\frac{1}{4}+E_i +|s n_i|^2}$ and $C=
1+m_i-s n_i$.} this 
happens when
\bea
\frac{1}{2}-\sqrt{\frac{1}{4}+E_i +|s n_i|^2} +
\mbox{max} ( |sn_i|, |m_i| ) = 0, -1, -2, \cdots.
\eea
By writing $l_i=k_i + \mbox{max} (|sn_i|, |m_i|) $ where $k_i =0, 1, 2,
\cdots$ we get $E_i=l_i(l_i+1)-n_i^2 s^2$.
Therefore the 
eigenvalue of the Laplacian on $Q^{n_1, n_2, n_3}$ from (\ref{eigen})
is given by
\bea
E= \sum_{i=1}^3 \La_i \left( l_i(l_i+1)-n_i^2 s^2 \right) +\frac{s^2}{c^2}
\label{Eigen}
\eea
where $l_i=|n_i s|, |n_i s|+1, \cdots $.
The eigenvalue $E$ is 
classified by $U(1)$ charge $s$ and spins $l_i$'s under 
$SU(2) \times SU(2) \times SU(2)$. The eigenmodes occur in the
${\bf (2l_1+1, 2l_2+1, 2l_3+1)}$ dimensional representation of $SU(2)
\times SU(2) \times SU(2)$ with $U(1)$ charges $s=0, \pm \frac{1}{2},
\pm 1, \cdots$.
The eigenvalues (\ref{Eigen}) as a linear combination of the quadratic 
Casimirs for the symmetry group $SU(2) \times SU(2) \times SU(2) \times U(1)$
are the form for  a coset manifold \cite{pope} sometime ago.

The dimension of the scalar operator in terms of energy labels,
 in the dual SCFT corresponding
$AdS_4 \times Q^{1,1,1}$ is
\bea
\Delta = \frac{3}{2} + \frac{1}{2} \sqrt{1 + \frac{m^2}{4}} =
\frac{3}{2} + \frac{1}{2} \sqrt{45 + \frac{E}{4} -6 \sqrt{36 +E}} .
\label{delta}
\eea 
The first equation in (\ref{delta}) 
comes from the relation between the lowest energy
eigenvalue and the mass which appears in the $AdS_4$ wave equation.
The relation of d'Alembertian in $AdS_4$
to Casimir operator was obtained in \cite{nicolai,cfh}. For the scalars,
$\Delta = \frac{3}{2} + \frac{1}{2} \sqrt{1 + m^2/4}$. 
The second relation in (\ref{delta}) comes from the formula of mass 
$m^2= E +176 -24 \sqrt{E +36}$ in \cite{df} where 
the normalization for this is
$(\Box -32 +m^2) S =0$ for scalar field $S$.
This can be read off directly from the eigenvalues of $\Box$ because
the relating fields have a mode expansion in terms of the scalar eigenfunctions
on $Q^{n_1, n_2, n_3}$. Other bosonic spectrum and fermionic spectrum can be
obtained by calculating higher spin Laplacian also which are more complicated.
See also \cite{under}.
Although the spectrum of dimensions on all the $Q^{n_1, n_2, n_3}$
is not much interested in,  $Q^{1, 1, 1}$ exhibits an interesting feature
which is relevant to superconformal algebra. In this case,
the Einstein condition implies that
$\La_i=4 \La/3$ and $1/c^2=8 \La/3$(coming from the 
explicit form of Ricci tensor \cite{pp,pp1} 
which we did not write down) and hence
from (\ref{Eigen})
\bea
E(Q^{1,1,1})=  \frac{4}{3} \La 
\left( \sum_{i=1}^3 l_i(l_i+1)-s^2 \right)
\eea
where $l_i \geq |s|, s=0, \pm \frac{1}{2}, \pm 1, \cdots$.
The $U(1)$ part of the isometry goup of $Q^{1, 1, 1}$
acts by shifting $s$. The integer $R$ charge, $R$ is related to $s$
by $s=R/2$. 
Let us take $R \geq 0$.  One can find the lowest value of $\Delta$ is
$R$ and corresponds to a mode scalar with $l_i=s$ because $E(Q^{1,1,1})$ 
becomes 
$32(2 s^2 +3s )$ with $\La=24$ and plugging back to (\ref{delta})
then $\De=R$. 
Thus we find
a set of operators filling out a ${\bf (R+1, R+1, R+1)_{R/2}}$ multiplet of
$SU(2) \times SU(2) \times SU(2) \times U(1)$ where a 
subscript is $U(1)$ charge
$R/2$ and the number ${\bf R+1}$ is 
the dimension of each $SU(2)$ representation.
{\it $\Delta=R$ saturates the bound on $\Delta$ from superconformal algebra}.
It was shown in \cite{fabbrietal} recently that 
from the harmonic analysis on $Q^{1,1,1}$ and the spectrum of 
$SU(2) \times SU(2) \times SU(2)$ representation of the $OSp(2|4)$ 
hypermultiplets, the hypermultiplet of conformal dimension $\De=R$ and
$U(1)$ charge $s=R/2$ should be in the representation $l_i=s=R/2$. 
\footnote{As pointed out in \cite{kw1}, there is a subtlety for finding the 
correct conformal dimension among two roots, $\De_{+}$  which we denoted as
simply $\Delta$ in (\ref{delta}) and $\De_{-}$ with minus sign, in the 
supergravity side. So far we assumed that the integer number $R$ is greater 
than 1. So the square root of first equation in (\ref{delta}) gives rise to
$2R-3$. However, when $R=1$, this expression goes like $3-2R$. Therefore,
we have to choose $\De_{-}$ in order to get the correct conformal dimension 
which is equal to $R$. So in our case, this is another example of the AdS/CFT
duality where the unconventional $\De_{-}$ branch has to be chosen for the
operators which has $R$ charge of 1. Recall that the $\S^7$ case where 
the spherical harmonics correspond to traceless symmetric tensors of $SO(8)$.
All chiral operators in the ${\cal N}=8$ $SU(N)$ theory correspond to the 
conventional branch of dimension $\De_{+}$ except just one case. It is well 
known that this family of operators with dimension $\De=k/2, 
k=2, 3, \cdots$ is $\mbox{Tr} X^{i_1} X^{i_2} \cdots X^{i_k}$ where 
$X^i$ are the scalars in the vector multiplet. Using $\De(\De-3)=m^2$, it is
easy to see that $\De=k/2=\De_{+}, k=3, 4, \cdots$ and $\De=k/2=\De_{-}, 
k=2$.}

According to \cite{fabbri1}, 
the information on the Laplacian eigenvalues allows us 
to get the spectrum of hypermultiplets of the theory corresponding to
the chiral operators of the CFT.
This part of spectrum was given in \cite{fabbrietal} and the form of operators
is $\mbox{Tr} (ABC)^R$ where the $SU(2)$'s indices are totally symmetrized. 
From this, the dimension of $ABC$ should be 1.
Although the complete KK spectrum is not known yet,
we expect that the relevant operators in higher towers are descendant fields of
$\mbox{Tr} (ABC)^R$ which would have the form of 
$\mbox{Tr} F^2_1 (ABC)^R + \mbox{Tr} F^2_2 (BCA)^R + \mbox{Tr} F^2_3 (CAB)^R$.
Although the dimension of nonchiral operators are in general irrational,
there exist special integer values of $k_i$ such that
for $l_i=k_i+s$, the Diophantine like condition, 
$-2(k_1 k_2+k_2 k_3+k_3 k_1)+\sum_{i=1}^3 (k_i^2- k_i)=0$ 
make $\sqrt{36+E}$ be equal to
$8s+2(2\sum_{i=1}^3 k_i +3)$. Furthermore in order to make 
the dimension be rational, $45 + E/4 -6 \sqrt{36 +E}$ should be square of 
something. It turns out this is the case without any further restrictions on
$k_i$'s. Therefore we  have $\De=R+\sum_i^3 k_i$ which is $\De_{+}$ for
$\De \geq 3/2$ and $\De_{-}$ for $\De \leq 3/2$. 
Now we list some operators whose conformal dimensions are integers 
in terms of their representation $\bf (2l_1+1, 2l_2+1, 2l_3+1)_s$.
\bea
& & \De_{-}=1:  {\bf 
(1, 1, 3)_0, \;\; (1, 3, 1)_0, \;\; (3, 1, 1)_0, \;\; (2, 2, 2)_
{\pm 1/2} },\nonu \\
& & \De_{+}=2: { \bf 
(2, 2, 4)_{\pm1/2}, \;\; (2, 4, 2)_{\pm 1/2}, \;\; (4, 2, 2)_{
\pm 1/2} }, \;\; {\bf (3, 3, 3)_{\pm 1}}, \nonu \\
& & \De_{+} =3: {\bf (3, 3, 5)_{\pm 1}}, \;\; {\bf (3, 5, 3)_{\pm 1}}, \;\;
{\bf (5, 3, 3)_{\pm 1}}, \;\; {\bf (4, 4, 4)_{\pm 3/2}}.
\eea
From the discussion of \cite{pope}, the first series($\De=1$) 
give rise to possess 
extra massless supermultiplets while from the second series($\De=2$) there
are additional massless $0^{+}$ in massive supermultiplets.
The supermultiplet containing $\bf (2, 2, 2)_{\pm 1/2}$ has to include 
another scalars and one of them corresponds to the lower component of
the superfield $\mbox{Tr} (ABC)$ which has dimension $\De_{-}=1$ while other 
corresponds to the upper component which has dimension $\De_{+}=2$. Therefore 
supersymmetry requires that one chooses dimension $\De_{+}$ for one scalar
and $\De_{-}$ for the other.
It is easy to check that the value of $45 + E/4 -6 \sqrt{36 +E}$ is greater
than equal to 0 for all possible values of $E$. There are no states below the
Breitenlohner-Freedman bound \cite{bf}. It is possible to have the second 
solution with minus sign in front of square root in (\ref{delta}) provided
the conformal dimension $\De$ is greater than or equal to $1/2$ which is
a unitary bound. Whether decendant fields whose dimensions are larger than the
dimension of its chiral primary parent are protected or not will be clear when
one understands the full supergravity solution.  


\section{Baryon like Operators}

By putting a large number of $N$ of coincident M2 branes at the
conifold singularity and taking the near horizon limit,
the metric becomes  that \cite{ffgt,fabbrietal} of $AdS_4 \times Q^{1,1,1}$
\bea
ds_{11}^2 = \frac{r^4}{L^{2/3}} \eta_{\mu \nu} d y^{\mu} d y^{\nu} +
L^{1/3} \left(  \frac{dr^2}{r^2} +   g_{ij} dx^i dx^j \right).
\eea
The scale $L$ is related to $N$ by \cite{fabbrietal}
\bea
L= \left(\frac{\La}{6} \right)^{-3} =
\ell_p^6 2^5 \pi^2 N \frac{\mbox{Vol}(\S^7)}{\mbox{Vol}(Q^{1,1,1})}
\label{scale}
\eea 
where $\ell_p$ is a Planck scale which is the only universal parameter
in M theory and $\mbox{Vol}(\S^7)= \pi^4(6/\La)^{7/2}/3$. 
\footnote{The normalization \cite{fabbrietal} 
for four-form field strength is 
$G_{ijkl} = e \ep_{ijkl}$ where the parameter $e$ is a real constant. By 
plugging this into the 11 dimensional field equations, it leads to the 
product of 4 dimensional Einstein space, $R_{\mu \nu}= -
2 \La \eta_{\mu \nu}$ with Minkowski signature$(-,+,+,+)$ and 7 dimensional 
Einstein space $R_{ij}= \La g_{ij}$ where $\La$ is defined by $\La=
24 e^2/\kappa^{4/9}$ through grvitational constant $\kappa$. Moreover 
$\kappa^2= 8 \pi G_{11} = (2\pi)^8 \ell_p^9/2$.} 
The first equation arises when we write $AdS_4$ radius 
in terms of both cosmological constant $\La$ and scale factor $L$.
Since M2 branes
have the operators with dimension $\sqrt{N}$ by M2 tension formula
and M5 branes have the 
operators with dimension $N$ through the relation between mass, tension
\cite{dealwis}
and volume of branes,
we consider wrapping a M5 brane over 5-cycle of $Q^{1,1,1}$. Three 
5-cycles spanning $H_5(Q^{1,1,1})$ are the restrictions of the 
$U(1)$ fibration to the product of two of the three ${\bf P}^1$'s. A 5-cycle 
of minimum volume is to take the subspace at a constant value of 
$(\theta_3, \phi_3)$ in the metric (\ref{metric1}). To calculate the 5 volume,
$\mbox{Vol(5-cycle)}$,
it is necessary to find the determinant of the following metric by taking
the subspace at a constant value of, for example,  
$(\theta_3, \phi_3)$ in the metric (\ref{metric1})
\bea
\frac{3}{8 \La} \left(d\ps +\sum_{i=1}^{2} \cos \th_i d \ph_i \right)^2
 + \frac{3}{4 \La} \sum_{i=1}^{2}  \left( d \th_i^2+
\sin^2 \th_i d \ph_i^2 \right).
\eea  
By integrating the square root of the determinant over the five coordinates,
one can find 
\bea
\mbox{Vol(5-cycle)} = \frac{\pi^3}{4} \left( \frac{6}{\La} \right)^{5/2}.
\eea
Other two 5-cycles can be obtained by changing the role of three 
${\bf P}^1$'s and it turns out that 
their volumes are the same.
The mass of the M5 brane wrapped over 5-cycle, given by M5 brane tension times
$\mbox{Vol(5-cycle)}$, is
\bea
m= \frac{1}{(2 \pi)^5 \ell_p^6} \mbox{Vol(5-cycle)}.
\label{mass}
\eea
By the relation (\ref{delta})    
\bea
m^2= \frac{2\La}{3} (\De-1)(\De-2) \approx  \frac{2\La}{3} \De^2
\eea
for large $\De$ and
the relations (\ref{mass}) and (\ref{scale}),
one  gets for the mass formula \cite{fabbrietal} for the dimension of a baryon
corresponding to the M5 brane wrapped 5-cycle
\bea
\De= \frac{\pi N}{\La} \frac{\mbox{Vol(5-cycle)}}{\mbox{Vol}(Q^{1,1,1})}=
\frac{N}{3}
\label{nover3}
\eea
where the volume of $Q^{1,1,1}$ is $\mbox{Vol}(Q^{1,1,1})=\frac{\pi^4}{8} 
(\frac{6}{\La})^{7/2}$, given by (\ref{volumeQ}).

Next thing we do is to find corresponding operators in dual field theory
whose dimension is $N$.
Since the fields $A^{\al}_{k \be}$ carry an index $\al$ in the $\bf N$ of 
$SU(N)_1$ and an index $\be$ in the $\bf {\overline{N}}$ of $SU(N)_2$,
one can construct a baryon like color singlet operator by antisymmetrizing
completely with repect to both groups. The resulting gauge invariant 
chiral operator is
\bea
{\cal B}_{1l} = \ep_{\al_1 \cdots \al_N} \ep^{\be_1 \cdots \be_N} 
D_l^{k_1 \cdots k_N} \prod_
{i=1}^N A^{\al_i}_{k_i \be_i}
\label{b1}
\eea 
where $ D_l^{k_1 \cdots k_N}$ is the completely symmetric $SU(2)$
Clebsch-Gordon coefficient corresponding to forming the $\bf N+1$ of $SU(2)$
out of $N$ $\bf 2$'s. Therefore, the $SU(2) \times SU(2) \times SU(2)$
quantum numbers of ${\cal B}_{1l}$ are $\bf (N+1, 1, 1)$. 
In order to understand these $SU(2)$ quantum numbers, it is necessary to
do collective coordinate quantization of the wrapped M5 brane along the
five coordinates $(\psi, \th_2, \ph_2, \th_3, \ph_3)$ which acts as
a charged particle because nonzero Betti number of $Q^{1,1,1}$ implies 
nonperturbative states only which can be charged. 
Since the lowest angular momentum of
a charge particle is $N/2$, the ground state collective coordinate wave 
functions form a $\bf N+1$ 
dimensional representation of the first $SU(2)$ which
rotates the first $\S^2$. This $\S^2$ is not wrapped by M5 brane because
it is localized at a constant coordinate $(\th_1, \ph_1)$. 
Of course, wrapped
M5 brane is a singlet under other $SU(2)$'s.

Similarly, one
can construct baryon like operators which transform as $\bf (1, N+1, 1)$,
\bea
{\cal B}_{2l} = \ep_{\be_1 \cdots \be_N} \ep^{\ga_1 \cdots \ga_N} 
D_l^{k_1 \cdots k_N} \prod_
{i=1}^N B^{\be_i}_{k_i \ga_i}.
\label{b2}
\eea
The $SU(2) \times SU(2) \times SU(2)$
quantum numbers of ${\cal B}_{2l}$ in this case are $\bf (1, N+1, 1)$. 
The fields $B^{\be}_{k \ga}$ carry an index $\be$ in the $\bf N$ of 
$SU(N)_2$ and an index $\ga$ in the $\bf {\overline{N}}$ of $SU(N)_3$.
The ground state collective coordinate wave 
functions represent a $\bf N+1$ 
dimensional representation of the second $SU(2)$ which
rotates the second $\S^2$.
Finally 
baryon like operators which transform as $\bf (1, 1, N+1)$ are
\bea
{\cal B}_{3l} = \ep_{\ga_1 \cdots \ga_N} \ep^{\al_1 \cdots \al_N} 
D_l^{k_1 \cdots k_N} \prod_
{i=1}^N C^{\ga_i}_{k_i \al_i}.
\label{b3}
\eea
Here 
the fields $C^{\ga}_{k \al}$ carry an index $\ga$ in the $\bf N$ of 
$SU(N)_3$ and an index $\al$ in the $\bf {\overline{N}}$ of $SU(N)_1$.
The ground state collective coordinate wave 
functions represent a $\bf N+1$ 
dimensional representation of the third $SU(2)$ which
rotates the third $\S^2$.
Under the symmetry which exchanges the fundamental fields $A, B, C$ of the 
gauge groups $SU(N)_1 \times SU(N)_2 \times SU(N)_3$, these operators map
to M5 branes localized at either constant $(\th_2, \ph_2)$ or 
$(\th_3, \ph_3)$. The existence of three types of baryon operators is related 
to the fact that the base of $U(1)$ bundle of internal space is $ 
\S^2 \times \S^2 \times \S^2$.
Since each of $A, B, C$ has dimension $1/3$ in the construction 
\cite{fabbrietal} due to the fact that we have seen the conformal 
dimension of $ABC$ is 1 in the previous section and a permutation symmetry 
among them, 
the dimension of the baryon like operators is $N/3$ which 
is in agreememt with supergravity calculation we have worked before 
(\ref{nover3}). This implies that {\it three 5-cycles are supergravity
representations of conformal operators 
(\ref{b1}), (\ref{b2}) and (\ref{b3})}.

For consistency check, one can consider the dimension of Pfaffian operator
in $SO(2N)$ gauge theory. Gauge invariant baryonic operator 
$\ep_{a_1 \cdots a_{2N}} \Phi^{a_1 a_2} \cdots \Phi^{a_{2N-1} a_{2N}}$ has 
dimension $N/2$. The $SO(2N)$ theory is dual to $AdS_4 \times \RP^7$ and
the dual Pfaffian wrapping M5 brane on a $\RP^5$ gives, according to
(\ref{nover3})
\bea
\De= \frac{\pi N}{\La} \frac{\mbox{Vol(5-cycle)}}{\mbox{Vol}(X_7=\RP^7)} 
 =\frac{\pi N}{\La} \frac{\mbox{Vol}(\RP^5)}{\mbox{Vol}(\RP^7)}=\frac{N}{
2}.
\eea
Moreover, an ${\cal N}=2$ theory \cite{aot} 
results from $\Z_3$ orbifold action on $\S^7$ 
defined by coordinatizing $\R^6$ by three complex numbers $z_1, z_2, z_3$
orthogonal to M2 brane worldvolume and considering the map 
$z_k \rightarrow e^{2\pi i/3} z_k$ for all $k$. Minimal area 5-cycles on
$\S^7/\Z_3$ can be constructed by intersecting the 5-plane $z_k=0$ for any
particular $k$ with the sphere $|z_1|^2+|z_2|^2+|z_3|^2=1$. Then
\bea
\De= \frac{\pi N}{\La} \frac{\mbox{Vol(5-cycle)}}{\mbox{Vol}(X_7=\S^7/\Z_3)}=
 \frac{\pi N}{\La} \frac{\mbox{Vol}(\S^5)/3}{\mbox{Vol}(\S^7)/3}=\frac{N}{2}.
\eea
The theory has gauge group $SU(N)_1 \times SU(N)_2 \times SU(N)_3$ with 
three $\bf (N, \overline{N})$ representations between each pair of gauge 
groups. Baryon like operators as in (\ref{b1}), (\ref{b2}) or (\ref{b3})
from the bifundamental matter have dimension $N/2$.
  
\section{Domain Walls in $AdS_4$ and Other Wrapped Branes}

Since $AdS_4$ has three spatial dimensions, M2 branes in $AdS_4$ could 
potentially behave as a domain wall. Since M2 brane is the electric source of
the four form field $G_4$, 
the integrated hodge dual of 
four-form flux over 7 manifold $\int_{X_7} \star G_4 $ jumps
by one unit when one crosses the domain wall. This means the gauge group
of the boundary conformal field theory can change, for example, 
from $SU(N)_1 \times SU(N)_2 \times SU(N)_3$ on one side to
$SU(N+1)_1 \times SU(N+1)_2 \times SU(N+1)_3$ on the other side for
$AdS_4 \times Q^{1,1,1}$. Of course, for the anti-M2 brane, the gauge group 
will change $SU(N-1)_1 \times SU(N-1)_2 \times SU(N-1)_3$. 
The similar situation also occurs when 
M5 brane is wrapped on a specific 3-cycle of $Q^{1,1,1}$ to make a M2 brane in
$AdS_4$. 
Using the orthonormal bases generated by the vielbeins of $Q^{1,1,1}$ for
given metric (\ref{metric1})
\bea
e^{\psi}=
\sqrt{\frac{3}{8 \La}} \left(d\ps +\sum_{i=1}^{3} \cos \th_i d \ph_i \right),
\;\;\; e^{\th_i}=
  \sqrt{\frac{3}{4 \La}} d \th_i,
\;\;\; e^{\phi_i}= \sqrt{\frac{3}{4 \La}} \sin \th_i d \ph_i
\eea
where $i=1, 2, 3$,
the harmonic representatives of second, third and fifth cohomology groups
can be written in terms of these combinations.
Note that from 4-th cohomology, $H^4(Q^{1,1,1}, {\bf Z})={\bf Z}_2 \cdot
(\omega_1 \omega_2+\omega_1 \omega_3+\omega_2 \omega_3)$ where $\omega_i$'s 
are the generators of the second cohomology group of the ${\bf P}^1$'s, 
3rd homology $H_3(Q^{1,1,1},\Z)$ can be obtained. 
In general, 3-cycle can be viewed as a fibration of $\ps$ over 2 sphere
parametrized by some combination of $(\th_i, \ph_i), i=1, 2, 3$, but
we are thinking a domain wall(a M5 brane wrapped over 3-cycle) 
together with baryon like operator(a M5 brane wrapped over 5-cycle). In the
previous section, we considered 5-cycle as  a fibration of $\ps$ over  
$(\th_i, \ph_i, \th_j, \ph_j), i, j=1, 2, 3$. We take 3-cycle as 3 dimensional
space orthogonal to 5-cycle except one common direction.
There exist three ways for 
M5 brane to be wrapped on 3-cycle depending on the choice of three $\S^2$'s 
in the base of $U(1)$ bundle.
When on one side of the domain wall one has the original $SU(N)_1 
\times SU(N)_2 \times SU(N)_3$, then on the other side the corresponding one is
$SU(N)_1 \times SU(N)_2 \times SU(N+1)_3$ if we take first 3-cycle. 
The matter fields
$A_i, B_j, C_k$ are filling out $2 {\bf (N, \overline{N}, 1)} \oplus 2 {\bf
(1, N, \overline{N+1})} \oplus 2 {\bf (\overline{N}, 1, N+1)}$. 
Similarly, if we take second 3-cycle, 
on the other side the corresponding gauge
group is $SU(N)_1 \times SU(N+1)_2 \times SU(N)_3$. The matter fields 
$A_i, B_j, C_k$ are filling out  $2 {\bf (N, \overline{N+1}, 1) } 
\oplus 2 {\bf (1, N+1, \overline{N})} \oplus 2 {\bf 
(\overline{N}, 1, N)}$.      
Also if we take third 3-cycle, 
on the other side the corresponding gauge
group is $SU(N+1)_1 \times SU(N)_2 \times SU(N)_3$. The matter fields 
$A_i, B_j, C_k$ are filling out  $2 {\bf (N+1, \overline{N}, 1) } 
\oplus 2 {\bf (1, N, \overline{N})} \oplus 2 {\bf (\overline{N+1}, 1, N)}$. 

Let us consider what happens if the baryons(wrapped M5 branes over 5-cycle) 
cross  a domain wall(wrapped M5 brane over 3-cycle).
Let us wrap M5 brane around particular 5-cycle which is invariant both under
the group $SU(2)_B$ under which the fields $B_j$ transform and
the group $SU(2)_C$ under which the fields $C_k$ transform.
Then the corresponding state in the $SU(N)_1 \times SU(N)_2 
\times SU(N)_3$ field
theory is ${\cal B}_1$ of (\ref{b1}). In the 
$SU(N)_1 \times SU(N+1)_2 \times SU(N)_3$ theory, we have
\bea
\ep_{\al_1 \cdots \al_N} \ep^{\be_1 \cdots \be_{N+1}} \prod_
{i=1}^N A^{\al_i}_{ \be_i}
\;\;\;\;\; \mbox{or} \;\;\;\;\;
\ep_{\al_1 \cdots \al_N} \ep^{\be_1 \cdots \be_{N+1}} \prod_
{i=1}^{N+1} A^{\al_i}_{ \be_i}
\eea
where
$SU(2)$ indices are omitted. These are either a 
fundamental of $SU(N+1)_2$ or a fundamental of $SU(N)_1$.  
This is no longer a singlet because when one antisymmetrizes the color indices
on a product of $N$ or $N+1$ bifundamentals of $SU(N)_1 \times SU(N+1)_2 
\times SU(N)_3$ 
there exists one free index. If we have M wrapped M5 branes over 
3-cycle rather than a single wrapped M5 brane over 3-cycle, then according to
$M$ units of  flux, the gauge group will be  $SU(N)_1 \times SU(N+M)_2 \times
SU(N)_3$.

The wrapped M5 brane must have M2 brane attached to it. The new aspect of 
the domain wall is that M2 brane must stretch from it to wrapped M5 brane.
Recall \cite{creation1,creation2}
that two M5 branes with one common direction cross, a M2 brane 
stretched between them is created. By dimensional reduction to the Type IIA 
string theory one can find T dual version of Hanany Witten effect 
\cite{hw}:When a NS5 brane and a D5 brane sharing 
two common directions pass through
each other, a D3 brane must be created.
The action containing a Chern Simon term is proportional to
$\int G_4(\th_2, \ph_2, \th_3, \ph_3)  
\wedge B_2(y^0, \psi)$ indicating that $G_4$ acts as a souce of
$B_2$ where $G_4 =d C_3$ is four-form field in M theory and $B_2$ is a
RR $B$ field.
From the flux through the baryonic M5 brane along $(y^0, \psi, 
\th_2, \ph_2, \th_3, \ph_3 )$ in the presence of 
domain wall M5 brane along $(y^0, y^2, y^3, \psi, \th_1, \ph_1)$, 
{\it the net charge that couples to B field gives rise to  
a M2 brane along $(y^0, y^1, \psi)$ stretched between M5 branes is created}.
We can reduce to Type IIA string theory along the $\psi$, which is common
to all branes. Then D4 brane along $(y^0, \th_2, \ph_2, \th_3, \ph_3)$ passing
D4 brane along $(y^0, y^2, y^3, \th_1, \ph_1)$ creates a fundamental string
along $(y^0, y^1)$ direction.

Similarly, 
If we consider M5 brane around particular 5-cycle which 
is invariant both under
the group $SU(2)_A$ under which the fields $A_i$ transform and
the group $SU(2)_C$ under which the fields $C_k$ transform.
Then the corresponding state in the $SU(N)_1 \times SU(N)_2 
\times SU(N)_3$ field
theory is ${\cal B}_2$ of (\ref{b2}). Then the corresponding 
$SU(N)_1 \times SU(N)_2 \times SU(N+1)_3$ theory has
\bea 
\ep_{\be_1 \cdots \be_N} \ep^{\ga_1 \cdots \ga_{N+1}} \prod_
{i=1}^{N} B^{\be_i}_{ \ga_i} \;\;\;\;\; \mbox{or}
\;\;\;\;\;
\ep_{\be_1 \cdots \be_N} \ep^{\ga_1 \cdots \ga_{N+1}} \prod_
{i=1}^{N+1} B^{\be_i}_{ \ga_i}.
\eea
which become a non singlet  because when one antisymmetrizes the color indices
on a product of $N$ or $N+1$ bifundamentals of $SU(N)_1 \times SU(N)_2 \times
SU(N+1)_3$ there exists one free index. That is, 
either  a 
fundamental of $SU(N+1)_3$ or a fundamental of $SU(N)_2$.  
Moreover, when we consider M5 brane over 5-cycle which 
is invariant both under
the group $SU(2)_A$ under which the fields $A_i$ transform and
the group $SU(2)_B$ under which the fields $B_j$ transform,
 the $SU(N)_1 \times SU(N)_2 \times SU(N)_3$ field
theory has ${\cal B}_3$ of (\ref{b3}). 
The corresponding 
$SU(N+1)_1 \times SU(N)_2 \times SU(N)_3$ theory has
\bea
\ep_{\ga_1 \cdots \ga_N} \ep^{\al_1 \cdots \al_{N+1}}  \prod_
{i=1}^N C^{\ga_i}_{ \al_i} \;\;\;\;\; \mbox{or} \;\;\;\;\;
\ep_{\ga_1 \cdots \ga_N} \ep^{\al_1 \cdots \al_N}  \prod_
{i=1}^{N+1} C^{\ga_i}_{ \al_i}.
\eea
In this case also there is one free index  
when one antisymmetrizes the color indices
on a product of $N$ or $N+1$ bifundamentals of $SU(N+1)_1 \times SU(N)_2 \times
SU(N)_3$.



From $H^3(Q^{1,1,1}, \Z)=H^6(Q^{1,1,1}, \Z)=0$,
there are no states associated with branes wrapping 4-cycle or 1-cycle.
For M5 branes, there are three types of wrapping because 
there exist nonzero $H_5(Q^{1,1,1},\Z), H_3(Q^{1,1,1}, \Z)$ and 
$H_2(Q^{1,1,1}, \Z)$. The first case involves 5-cycle and produces particle
in $AdS_4$ associated with baryon like operators (\ref{b1}), (\ref{b2}) and
(\ref{b3}). The second case involves 3-cycle and produces a domain wall we
have discussed. The last one involves M5 brane wrapping
2-cycle and produces threebrane in $AdS_4$. 
At this moment, we do not know how this can be realized in the full M theory 
and it is not clear what is interpretation of boundary conformal field theory. 
When M5 brane is wrapped around
the eleventh circle $\S^1$ orthogonal to 2-cycle, then in Type 
IIA description, this is equivalent to  twobrane in $AdS_4$.    
For M2 branes, there exists only one type of wrapping which involves
2-cycle and produces particle in $AdS_4$ because M2 brane can not wrap
higher dimensional space 5-cycle or 3-cycle. The mass of M2 brane wrapped
2-cycle goes like $\sqrt{N}$ from the mass formula which was not appropriate 
for the candidate of baryon like operator that must behave like as $N$.  

\section{Conclusion}

In summary, 
since $AdS_4 \times Q^{1,1,1}$ is a supersymmetric holographic theory based
on a compact manifold $Q^{1,1,1}$ which is  not locally $\S^7$, the dual
$SU(N)_1 \times SU(N)_2 \times SU(N)_3$ gauge theory can not be obtained from
a projection of the ${\cal N}=8$ theory. The dual representation of
baryon like operators from a symmetric product of $N$ bifundamental matter
fields, fully antisymmetrized on upper and lower indices separately,
is a M5 brane wrapped around an 5-cycle in $Q^{1,1,1}$. Three ways of 
embedding 5-cycle are orbits of two of the three $SU(2)$ global symmetry
groups of the theory. A M5 brane wrapping 5-cycle can be regarded as
a charge particle allowed to move on the 2-cycle parametrizing 
remaining orbits. The 5 volume of the $SU(2)$ orbits gives a dimension for
the operators $N/3$ which matches 
exactly the field theory. By using the
baryon like operators, wrapped M5 branes around 3-cycle of
$Q^{1,1,1}$ behave as a domain wall separating the original 
$SU(N)_1 \times SU(N)_2 \times SU(N)_3$ from,  for example, 
$SU(N)_1 \times SU(N+1)_2 \times SU(N)_3$. 
The crucial point was that a M2 brane is
created when a M5 brane(wrapped around 5-cycle) crosses a M5 brane(wrapped 
around 3-cycle). This means that a baryon is no longer a singlet that agrees 
with the field theory observation, because there exists one free index when 
antisymmetrizing the color indices on a product of $N$ or $N+1$ 
bifundamentals of $SU(N)_1 \times SU(N+1)_2 \times SU(N)_3$.

There are various $G/H$ models which has at 
least a supersymmetry less than or equal to three in 3 dimensions. It is 
already known that there exists a dual conformal field theory corresponding 
to $M^{pqr}$ space. It is natural to ask how other cases can be realized in 
the boundary conformal field theories. It would be interesting to study
whether one can find wrapping branes over various cycles of possible 
$X_7$'s and discuss their field theory interpretations.    
The theory of M2 branes at orbifold singularity, for example, $\C^4/\Ga$
tells us a variety of supersymmetric theories \cite{mp} 
depending on how $\Ga$ acts.
It is not known what is boundary conformal field theory corresponding to
M2 branes at $\C^4/(\Z_2 \times \Z_2)$ \cite{ahnkim}. It is quite 
interesting to see 
how our conifold description can be obtained from a deformation of
orbifold singularity.
      
\vskip1cm

We would like to 
thank K. Oh and R. Tatar for collaboration in the early stages of 
this project.

\end{document}